\documentclass{article}

\usepackage{arxiv}
\usepackage{multirow, multicol} 
\usepackage[utf8]{inputenc} 
\usepackage[T1]{fontenc}    
\usepackage{hyperref}       
\usepackage{url}            
\usepackage{booktabs}       
\usepackage{amsfonts}       
\usepackage{nicefrac}       
\usepackage{microtype}      
\usepackage{lipsum}
\usepackage{import}
\usepackage{soul,color}
\usepackage{array}

\usepackage{stackengine}
\usepackage{subcaption}
\usepackage{amssymb}
\usepackage{graphicx}
\usepackage{import}
\usepackage{comment}
\usepackage{soul,color}
\usepackage{amsmath}
\usepackage{multirow, multicol}
\usepackage{svg}
\usepackage[utf8]{inputenc} 
\usepackage[T1]{fontenc}    
\usepackage{hyperref}       
\usepackage{url}            
\usepackage{booktabs}       
\usepackage{amsfonts}       
\usepackage{nicefrac}       
\usepackage{microtype}      
\usepackage{lipsum}
\usepackage{import}
\usepackage{soul,color}
\usepackage{array}
\usepackage{footnote}

\RequirePackage{tikz} 

\newcommand{\orcidicon}{\includegraphics[width=0.32cm]{figures/definitions/logo-orcid.eps}}

\foreach \x in {A, ..., Z}{%
\expandafter\xdef\csname orcid\x\endcsname{\noexpand\href{https://orcid.org/\csname orcidauthor\x\endcsname}{\noexpand\orcidicon}}
}

\usepackage{longtable}
\usepackage[acronym]{glossaries}

\title{Music Composition with Deep Learning: A Review}

\author{
  Carlos Hernandez-Olivan \thanks{https://carlosholivan.github.io} \\
  \texttt{carloshero@unizar.es} \\
   \And
 Jose R. Beltran \\
  \texttt{jrbelbla@unizar.es} \\
}

\begin{document}
\maketitle

\begin{abstract}
Generating a complex work of art such as a musical composition requires exhibiting true creativity that depends on a variety of factors that are related to the hierarchy of musical language. Music generation have been faced with Algorithmic methods and recently, with Deep Learning models that are being used in other fields such as Computer Vision. In this paper we want to put into context the existing relationships between AI-based music composition models and human musical composition and creativity processes. We give an overview of the recent Deep Learning models for music composition and we compare these models to the music composition process from a theoretical point of view. We have tried to answer some of the most relevant open questions for this task by analyzing the ability of current Deep Learning models to generate music with creativity or the similarity between AI and human composition processes, among others.

\end{abstract}


\section{Introduction} \label{sec:introduction}

Music is generally defined as a succession of pitches or rhythms, or both, in some definite patterns \cite{definition}. Music composition (or generation) is the process of creating or writing a new piece of music. The music composition term can also refer to an original piece or work of music \cite{definition}. Music composition requires creativity which is the unique human capacity to understand and produce an indefinitely large number of sentences in a language, most of which have never been encountered or spoken before \cite{chomsky2009syntactic}. This is a very important aspect that needs to be taken into account when designing or proposing an AI-based music composition algorithm.

More specifically, music composition is an important topic in the Music Information Retrieval (MIR) field. It comprises subtasks such as melody generation, multi-track or multi-instrument generation, style transfer or harmonization. These aspects will be covered in this paper from the point of view of the multitude of techniques that have flourished in recent years based on AI and DL.

\subsection{From Algorithmic Composition to Deep Learning}
From the 1980s the interest in computer-based music composition has never stop to grow. Some experiments came up in the earlys 1980s such as the Experiments in Musical Intelligence (EMI) \cite{cope_emi} by David Cope from 1983 to 1989 or Analogiques A and B by Iannis Xenakis \cite{xenakis1981musiques}.
Later in the 2000s, also David Cope proposed the combination of Markov chains with grammars for automatic music composition and other relevant works such as Project1 (PR1) by Koening  \cite{Koening} were born.
These techniques can be grouped in the field of Algorithmic Music Composition, which is a way of composing by means of formalizable methods \cite{Todd1989} \cite{nierhaus2009algorithmic}. This type of composing consists on a controlled procedure which is based on mathematical instructions that must be followed in a fixed order. There are several methods inside the Algorithmic Composition such as Markov Models, Generative Grammars, Cellular Automata, Genetic Algorithms, Transition Networks or Caos Theory \cite{hiller1957musical}. Sometimes these techniques and other probabilistic methods are combined with Deep Neural Networks in order to condition them or help them to better model music, which is the case of DeepBach \cite{deepbach}. These models can generate and harmonize melodies in different styles, but the lack of generalizability capacity of these models and the rule-based definitions that must be done by hand make these methods less powerful and generalizable in comparison with Depp Learning-based models.

From the 1980s to the earlies 2000s, the first works which tried to model music with NNs were born \cite{Mozer94} \cite{Eck01} \cite{EckS02} \cite{bharucha1989modeling}. In the recent years, with the growing of Deep Learning (DL) lots of studies have tried to model music with deep Neural Networks (NN). DL models for music generation normally use NN architectures that are proven to perform well in other fields such as Computer Vision or Natural Language Processing (NLP). There can also be used pre-trained models in these fiels that can be used for music generation. This is called Transfer Learning \cite{ZhuangQDXZZXH21}. Some NN techniques and architectures will be shown later in this paper. Music composition today is taking input representations and NNs architectures from large-scale NLP applications, such as Transformer-based models, which are demonstrating very good performance in this task. This is due to the fact that music can be understood as a language in which every style or music genre has its own rules.

\subsection{Neural Network Architectures for Music Composition with Deep Learning}
First of all, we will provide an overview of the most widely used NN architectures that are providing the best results in the task of musical composition so far. The most used NN architectures in music composition task are Generative Models such as Variational AutoEencoders (VAEs) or Generative Adversarial Networks (GANs), and NLP-based models such as Long Short-Term Memory (LSTM) or Transformers. The following is an overview of these models.

\subsubsection{Variational Auto-Encoders (VAEs)}
The original VAE model \cite{vae} uses an Encoder-Decoder architecture to produce a latent space by reconstructing the input (see fig. \ref{fig:vae}). A latent space is a multidimensional space of compressed data in which the most similar elements are located closest to each other.
In a VAE, the encoder approximates the posterior and the decoder parameterizes the likelihood. The posterior and likelihood approximations are parametrized by a NN with $\lambda$ and $\theta$ parameters for the encoder and decoder respectively. The posterior inference is done by minimizing the Kullback-Leiber (KL) divergence between the encoder or approximate posterior, and the true posterior by maximizing the Evidence Lower bound (ELBO). The gradient is computed with the so-called \textit{reparametrization} trick. There are variations of the original VAE model such as the $\beta$-VAE \cite{beta_vae} which adds a penalty term $\beta$ to the reconstruction loss in order to improve the latent space distribution. In fig. \ref{fig:vae} we show the general VAE architecture. An example of a DL model for music composition based on a VAE is MusicVAE \cite{musicvae} which we describe in further sections in this paper.



\subsubsection{Generative Adversarial Networks (GANs)}
GANs \cite{gan} are Generative Models composed by two NNs: the Generator $G$ and the Discriminator $D$. The generator learns a distribution $p_g$ over the input data
The training is done in order to let the discriminator maximize the probability of assigning the correct label to the training samples and the samples generated by the generator. This training idea can be understood as if $D$ and $G$ follow the two-player minimax game that Goodfellow et al. \cite{gan} described. In fig. \ref{fig:gan} we show the general GAN architecture.


The generator and the discrimiator can be formed by different NN layers such as Multi-Layer Perceptrons (MLP) \cite{rosenblatt1958perceptron}, LSTM \cite{lstm} or Convolutional Neural Networks (CNN) \cite{FukushimaM82} \cite{CunHBB99}.
    
\subsubsection{Transformers}
\begin{figure}[h!]
    \begin{subfigure}[b]{1\textwidth}
    \centering
    \includegraphics[scale=0.18]{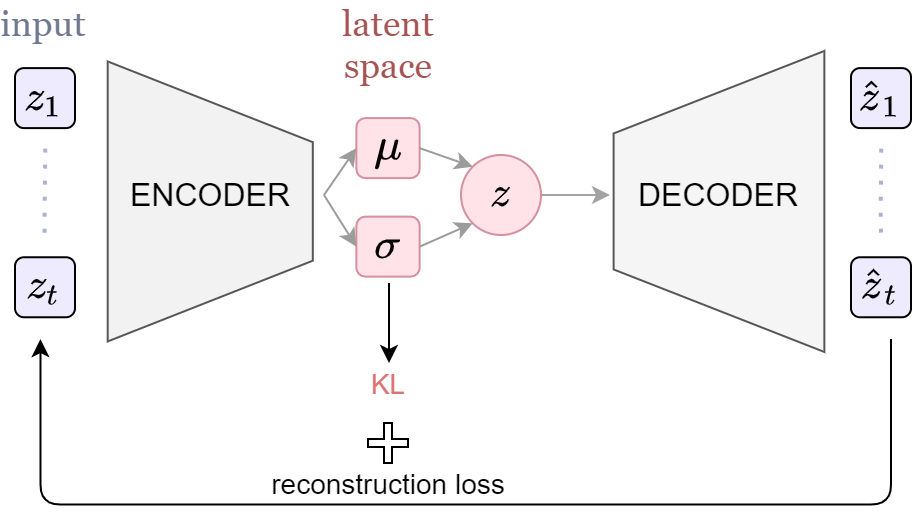}
    \caption{}
    \label{fig:vae}
    \end{subfigure}

    \begin{subfigure}[b]{1\textwidth}
        \centering
        \includegraphics[scale=0.15]{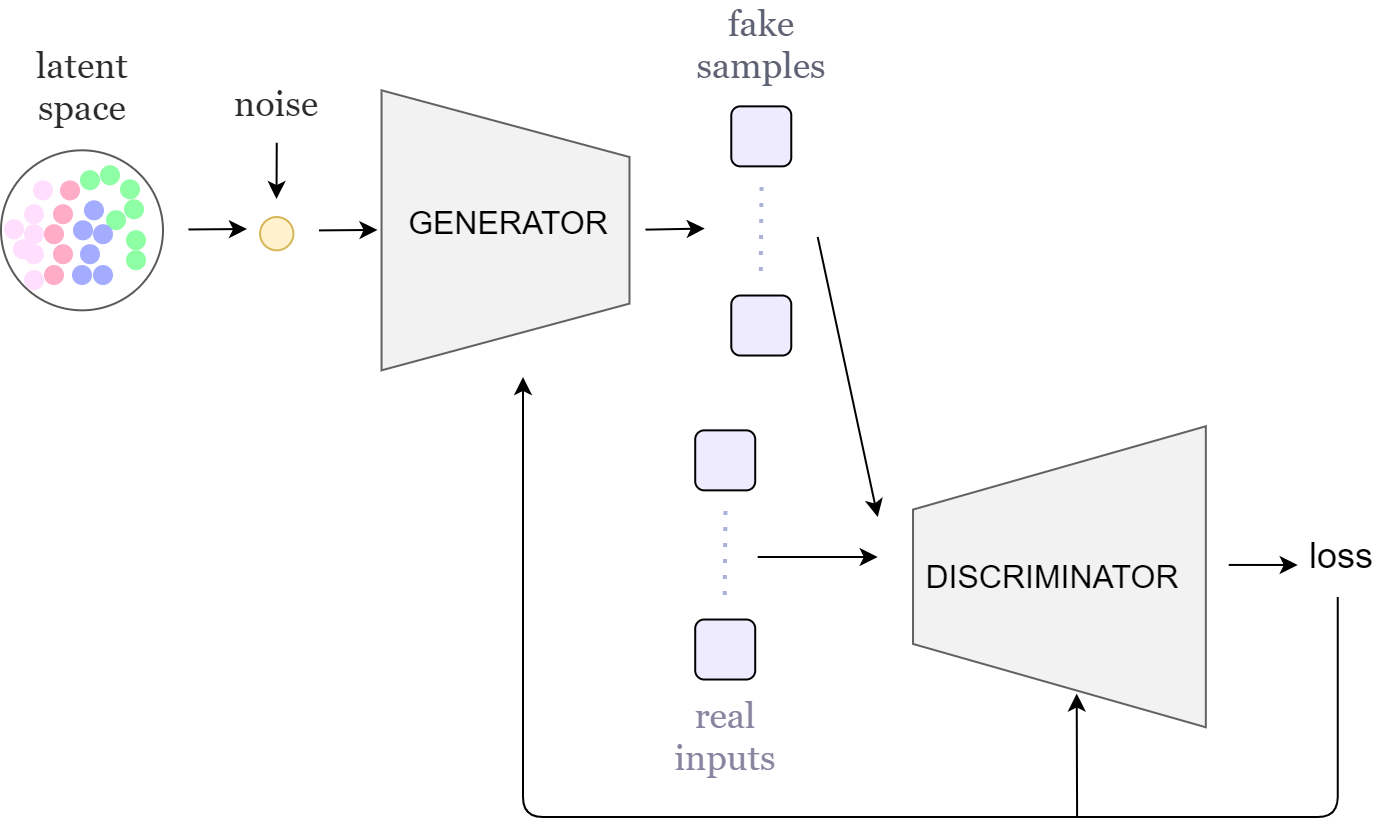}
        \caption{}
        \label{fig:gan}
    \end{subfigure}

    \begin{subfigure}[b]{1\textwidth}
        \centering
        \includegraphics[scale=0.14]{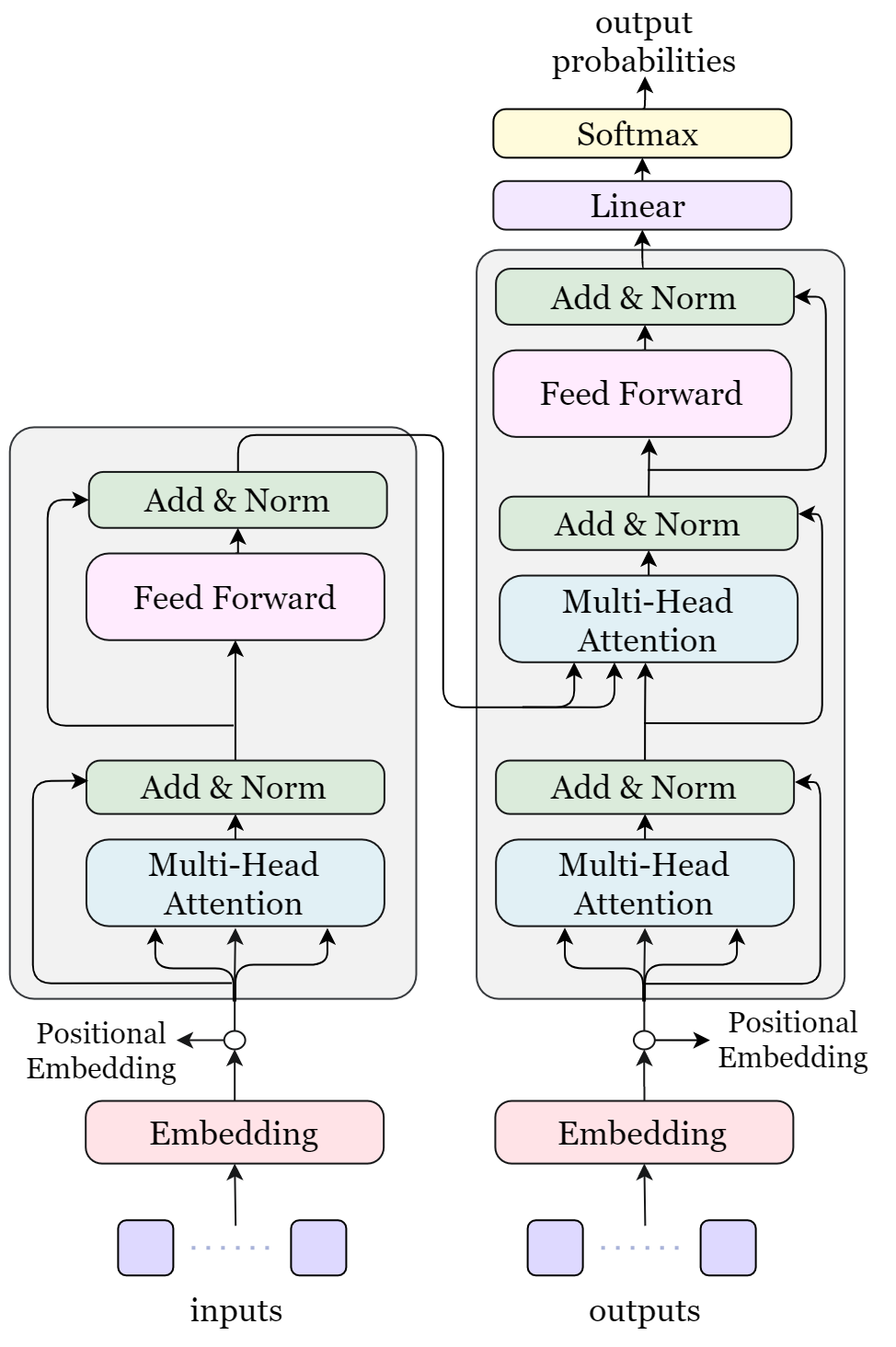}
        \caption{}
        \label{fig:transformer}
    \end{subfigure}
    \caption{a) VAE \cite{vae} b) GAN \cite{gan} and  c) Transformer general architecture. Reproduced from \cite{attention}.}
\end{figure}

Transformers \cite{attention} are being currently used in NLP applications due to their well performance not ony in NLP but also in Computer Vision models. Transformers can be used as auto-regressive models like the LSTMs which allow them to be used in generative tasks. The basic idea behind Transformers is the attention mechanism. There are several variations of the original attention mechanism proposed by Vaswani et al. \cite{attention} that have been used in the music composition task \cite{music_transformer}. The combination of the attention layer with feed forward layers leads to the formation of the Encoder and Decoder of the Transformer, which differs from purely AutoEncoder models that are also composed by the Encoder and Decoder. Transformers are trained with tokens which are structured representations of the inputs. In fig. \ref{fig:transformer} we show the general Transformer architecture.

\subsection{Challenges in Music Composition with Deep Learning} \label{sec:challenges}
There are different points of view from the challenges perspective in music composition with DL that make ourselves ask questions related to the input representations and DL models that have been used in this field, the output quality of the actual state-of-the-art methods or the way that researchers have measured the quality of the generated music. In this review, we ask ourselves the following questions that involve the composition process and output: Are the current DL models capable of generating music with a certain level of creativity? What is the best NN architecture to perform music composition with DL? Could end-to-end methods generate entire structured music pieces? Are the composed pieces with DL just an imitation of the inputs or can NNs generate new music in styles that are not present in the training data? Should NNs compose music by following the same logic and process as humans do? How much data do DL models for music generation need? Are current evaluation methods good enough to compare and measure the creativity of the composed music?


To answer these questions, we approach Music Composition or Generation from the point of view of the process followed to obtain the final composition and the output of DL models, i.e., the comparison between the human composition process and the Music Generation process with Deep Learning and the artistic and creative characteristics presented by the generated music. We also analyze recent state-of-the-art models of Music Composition with Deep Learning to show the result provided by these models (\textit{motifs}, complete compositions...). Another important aspect analyzed is the input representation that these models use to generate music to understand if these representations are suitable for composing music. This gives us some insights on how these models could be improved, if these Neural Network architectures are powerful enough to compose new music with a certain level of creativity and the directions and future work that should be done in Music Composition with Deep Learning.

\subsection{Paper Structure} \label{sec:paper}
In this review, we make an analysis of the Music Composition task from the composition process and the type of generated output, and we do not cover the performance or synthesis tasks. This paper is structured as follows. Section \ref{sec:process} introduces a general view of the music composition process and music basic principles. In Section \ref{sec:melody}, we give an overview of state-of-the-art methods from the melodic composition point of view and we describe the DL models that have been tested for composing structured music.  In Section \ref{sec:instrumentation} we describe DL models that generate multi-track or multi-instrument music, that is, music made for more than one instrument. In Section \ref{sec:evaluation} we show different methods and metrics that are commonly used to evaluate the output of a model for music generation. In Section \ref{sec:discussion} we describe the open questions that remain in music generation by analyzing the models which we describe in sections \ref{sec:melody} and \ref{sec:instrumentation}. Finally, in Section \ref{sec:future_work} we expose future work and challenges that are still open in research.

\section{The Music Composition Process} \label{sec:process}
As in written language, the music composition process is a complex process that depends on a large number of decisions \cite{levi1991field}. In the music field, this process \cite{collins2005synthesis} depends on the music style we are working with. As an example, it is very common in classical music to start with a small unit of one or two bars called \textit{motif} and develop it to compose a melody or music phrase, and in styles like pop or jazz it is more common to take a chord progression and compose or improvise a melody ahead of it. In spite of the music style we are composing in, when a composer starts a piece of music there is some basic melodic or harmonic idea behind it. From the classical music point of view, this idea (or \textit{motif}) is developed by the composer to construct the melody or phrase that follows a certain harmonic progression and then these phrases are structured in sections. Each section has its own purpose so it can be written in different keys and its phrases usually follow different harmonic progressions than the other sections. Normally, compositions have a melodic part and a accompaniment part. The melodic part of a piece of music can be played by different instruments whose frequency range may or may not be similar, and the harmonic part gives the piece a deep and structured feel. The instruments, that are not necessarily in the same frequency range, are combined with \textit{Instrumentation} and \textit{Orchestration} techniques (see section \ref{sec:structure}). These elements are crucial in musical composition and also an important key in defining the style or genre of a piece of music. Music, has two dimensions, the time and the harmony dimensions. The time dimension is represented by the notes duration or rhythm, which is the lowest level in this axis. In this dimension, notes can be grouped or measured in units called bars, that are ordered groups of notes. The other dimension, harmony, is related to the note values or pitch. If we think of an image, time dimension would be the horizontal axis and harmony dimension the vertical axis. Harmony does also have a temporal evolution but this is not represented in music scores. There is a very common software-based music representation called piano-roll that follows this logic.

The music time dimension  is structured in low-level units that are notes, which are grouped in bars that form (\textit{motifs}). In the time high-level dimension we can find sections which are composed by phrases that last eight or more bars (this depends on the style and composer). The harmony dimension lowest level is the note-level and then the superposition of notes also played by different instruments gives us chords. The sequence of chords are called chord progressions that are relevant to the composition and they also have dependencies in the time dimension. Being said that, we can think about music as a complex language model that consist on short and long-term relationships. These relationships extends in two dimensions, the time dimension which is related to music structure and the harmonic dimension which is related to the notes or pitches and chords, that is, the harmony. 

\begin{figure}[h]
     \centering
     \begin{subfigure}[b]{1\textwidth}
         \centering
         \includegraphics[width=0.85\linewidth]{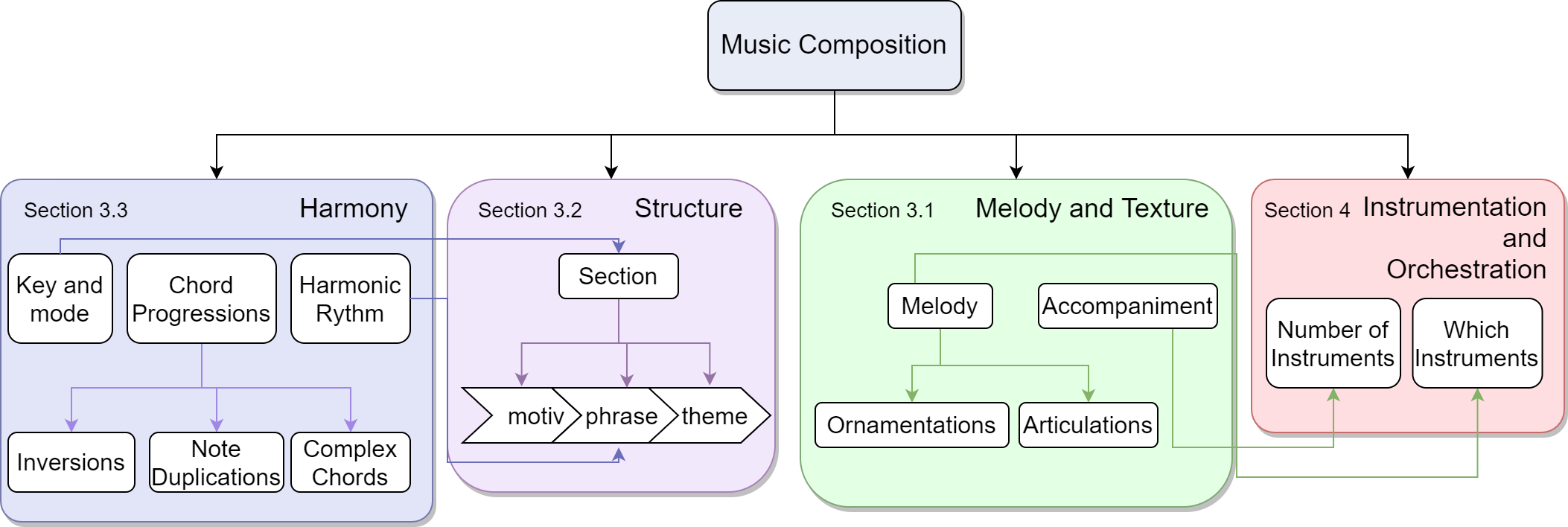}
         \caption{}
     \end{subfigure}

     \begin{subfigure}[b]{1\textwidth}
         \centering
         \includegraphics[width=0.7\linewidth]{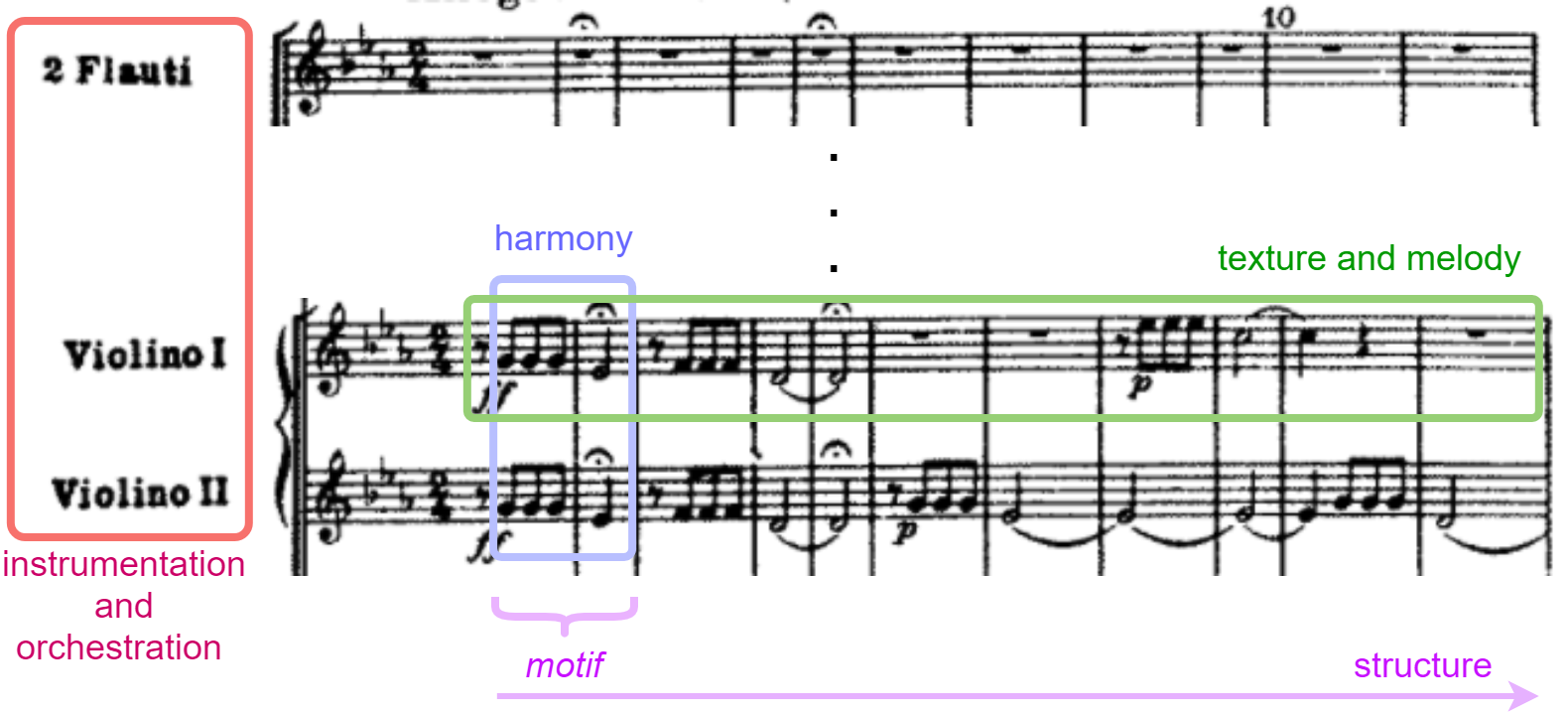}
    \caption{}
     \end{subfigure}
    \caption{a) General music composition scheme and b) an example of the beginning of Beethoven's 5th symphony with music levels or categories.}
    \label{fig:composing_scheme}
\end{figure}

From the symbolic music generation and analysis points of view, based on the ideas of Walton \cite{walton2005basic} the basic music principles or elements are (see fig. \ref{fig:composing_scheme}):
\begin{itemize}
    \item Harmony. It is the superposition of notes that form chords which compose a chord progression. The note-level could be considered as the lowest level in harmony being followed by the chord-level. The highest-level can be considered as the progression-level which usually belongs to a certain key.
    \item Music Form or Structure. It is the high-level structure that music presents and it is related with the time dimension. The smallest part of a music piece is the motif which is developed in a music phrase and the combination of music phrases form a section. Sections in music are ordered depending on the music style such as intro-verse-chorus-verse-outro for some pop songs (also represented as ABCBA) or exposition-development-recapitulation or ABA for Sonatas. The concatenation of sections which can be in different scales and modes gives us the entire composition.
    \item Melody and Texture. Texture in music terms refers to the melodic, rhythmic and harmonic contents that have to be combined in a composition in order to form the music piece. Music can be monophonic or polyphonic depending on the notes that are played at the same time step, homophonic or heterophonic depending on the melody, if it has or not accompaniment. 
    \item Instrumentation and Orchestration. These are music techniques that take into account the number of instruments or tracks in a music piece. Whereas Instrumentation is related to the combination of musical instruments which compose a music piece, Orchestration refers to the assignment of melodies and accompaniment to the different instruments that compose a determined music piece. In recording or software-based music representation, Instruments are organized as \textit{tracks}. Each track contains the collection of notes played on a single instrument \cite{ens2020mmm}. Therefore, we can call a piece with more than one instrument as multi-track, which refers to the information that contains two or more tracks where each track is played by a single instrument. Each track can contain one note or multiple notes that sounds simultaneously, leading to monophonic tracks and polyphonic tracks respectively.
\end{itemize}

Music categories are related between them. Harmony is related to the structure because a section is usually played in the same scale and mode. There are cadences between sections and there can also be modulations which change the scale of the piece. Texture and instrumentation are related to timbral features and their relationship is based on the fact that not all the instruments can play the same melodies. An example of that is when we have a melody with lots of ornamentation elements which cannot be played with determined instrument families (because of a fact of each instrument technique possibilities or a stylist reason). 


Another important music attribute are the dynamics, but they are related to the performance rather than the composition itself, so we will not cover them in this review. In Fig. \ref{fig:composing_scheme} we show the aspects of the music composition process that we cover in this review, and the relationships between categories and the sections of the paper in which each topic is discussed are depicted.

\section{Melody Generation} \label{sec:melody}

\begin{figure}[h]
    \centering
    \includegraphics[width=0.85\columnwidth]{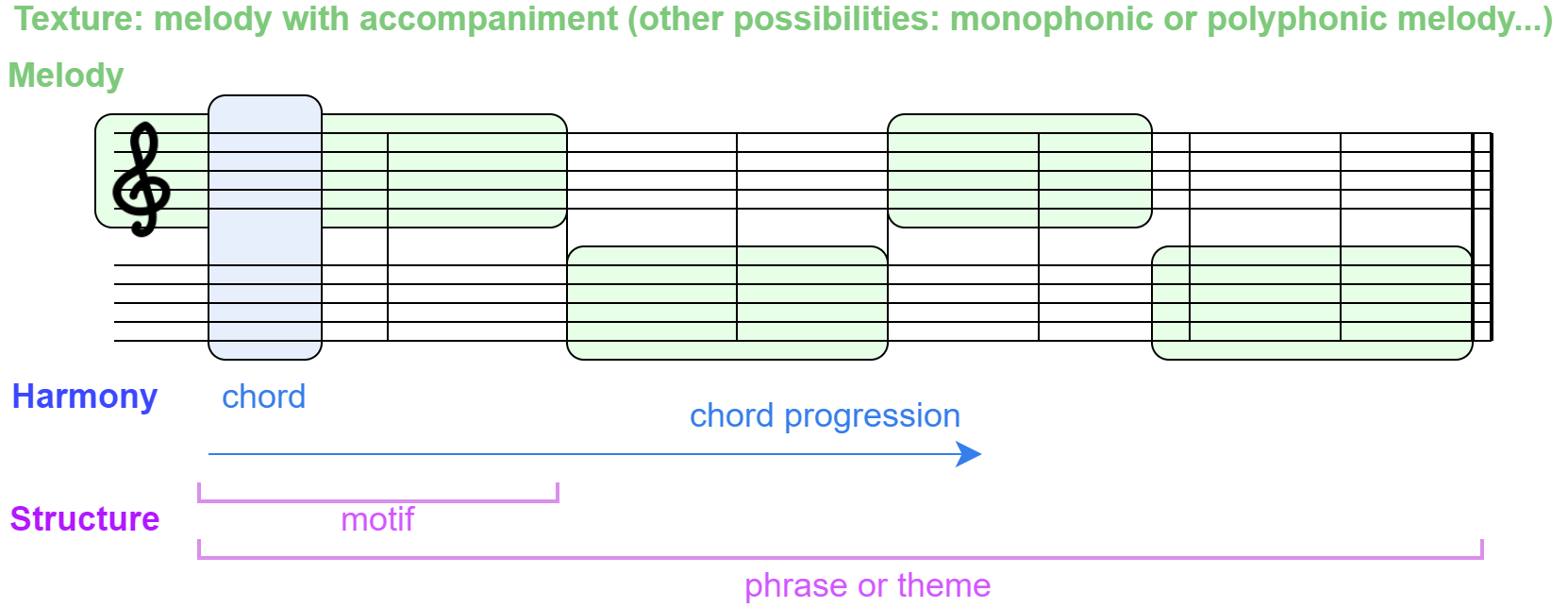}
    \caption{Scheme of an output-like score of melody generation models.}
    \label{fig:melody_scheme}
\end{figure}

A melody is a sequence of notes with a certain rhythm ordered in a aesthetic way. Melodies can be monophonic or polyphonic. Monophonic refers to melodies in which only one note is played at a time step whereas in polyphonic melodies there is more than one note being played at the same time step. Melody generation is an important part of music composition and it has been attempted with Algorithmic Composition and with several of the NN architectures that includes Generative Models such as VAEs or GANs, Recurrent Neural Networks (RNNs) used for auto-regression tasks such as LSTM, Neural Autoregressive Distriution Estimators (NADEs) \cite{nades} or current models used in Natural Language Processing like Transformers \cite{attention}. In Fig. \ref{fig:melody_scheme} we show the scheme with the music basic principles of an output-like score of a melody generation model.

\subsection{Deep Learning Models for Melody Generation: From Motifs to Melodic Phrases} \label{sec:sota}
Depending on the music genre of our domain, the human composition process usually begins with the creation of a \textit{motif} or a chord progression that is then expanded to a phrase or melody. When it comes to DL methods for music generation, several models can generate short-term notes sequences. In 2016, the very first DL models attempted to generate short melodies with Recurrent Neural Networks (RNNs) and semantic models such as Unit Selection \cite{unitselection}. These models worked for short sequences, so the interest to create entire melodies grew in parallel to the birth of new NNs. Derived from these first works and with the aim of creating longer sequences (or melodies), other models that combined NNs with probabilistic methods came up. An example of this is Google's Magenta Melody RNN models \cite{waite2016generating} released in 2016, the Anticipation-RNN \cite{abs-1709-06404} or DeepBach \cite{deepbach} both published in 2017. DeepBach is considered one of the current state-of-the-art models for music generation because of its capacity to generate 4-voices chorales in the style of Bach. 

However, these methods cannot generate new melodies with a high level of creativity from scratch. In order to improve the generation task, Generative Models were chosen by researchers to perform music composition. In fact, nowadays one of the best-performing models to generate \textit{motifs} or short melodies from 2 to 16 bars is MusicVAE\footnote{\url{https://magenta.tensorflow.org/music-vae}, accessed August 2021} \cite{musicvae} which was published in 2018. MusicVAE is a model for music generation based on a VAE \cite{vae}. With this model, music can be generated by interpolating in a latent space. This model is trained with approximately 1.5 million songs from the Lakh MIDI Dataset (LMD)\footnote{\url{https://colinraffel.com/projects/lmd/}, accessed August 2021} \cite{raffel2016learning} and it can generate polyphonic melodies for almost 3 instruments: melody, bass and drums. After the creation of MusicVAE model along with the birth of new NN architectures in other fields, the necessity and availability of new DL-based models that could create longer melodies grew. New models based on Transformers came up such as the Music Transformer \cite{music_transformer} in 2018, or models that used pre-trained Transformers such as the GPT-2 like MuseNet in 2019 proposed by OpenAI \cite{musenet} does. These Transformer-based models, such as Music Transformer, can generate longer melodies and continue a given sequence, but after a few bars or seconds the melody ends up being a bit random, that is, there are notes and harmonies that do not follow the musical sense of the piece. 

In order to overcome this problem and develop models that could generate longer sequences without losing the sense of the music generated in the previous bars or the main motifs, new models were born in 2020 and 2021 as combinations of VAEs, Transformers or other NNs or Machine Learning algorithms. Some examples of these models are the TransformerVAE \cite{transformer_vae} and PianoTree \cite{pianotree_vae}. These models perform well even in polyphonic music and they can generate music phrases. One of the latest released models to generate entire phrases is the model proposed in 2021 by Mittal et al. \cite{diffusion_models} which is based in Denoising Diffusion Probabilistic Models (DDPMs) \cite{ddpm} which are new Generative Models that generate high-quality samples by learning to invert a diffusion process from data to Gaussian noise. This model uses a MusicVAE 2-bar model to then train a Diffusion Model to capture the temporal relationships among the VAE latents $z_k$ with $k = 32$ which are the 32 latent variables that allows to generate 64 bars (2 bars per latent). In spite that there can be generated longer polyphonic melodies, they do not follow a central motif so they tend to loose the sense of a certain direction.

\subsection{Structure Awareness} \label{sec:structure}
As we mentioned in section \ref{sec:introduction}, music is a structured language. Once melodies have been created they must be grouped into bigger sections (see fig. \ref{fig:composing_scheme}) which play a fundamental role in a composition. These sections have different names that vary depending on the music style such as introduction, chorus or verse for pop or trap genres, and exposition, development or recapitulation for classical sonatas. Sections can also be named with capital letters and song structures can be expressed as ABAB, for example. Generating music with structure is one of the most difficult tasks in music composition with DL because structure means an aesthetical sense of rhythm, chord progressions and melodies that are concatenated with bridges and cadences \cite{Lattner_thesis}.

In DL there have been models that have tried to generate structured music by imposing the high-level structure with the Self-Similarity constrains. An example of that is the model proposed by Lattner et al. in 2018 \cite{Lattner_thesis} which uses a Convolutional Restricted Boltzmann Machine (C-RBM) to generate music and Self-Similarity constrain with a Self-Similarity Matrix \cite{muller2015fundamentals} to impose the structure of the piece as if it was a template. This method which imposes a structure template is similar to the composition process that a composer follows when composing music and the resulting music pieces followed the imposed structure template. Although new DL models are trending to be end-to-end, and new studies about modeling music with structure are being released \cite{chen2019effect}, there have not been DL models that could generate structured music by themselves, that is, without the help of a template or high-level structure information that is passed to the NN.

\subsection{Harmony and Melody Conditioning} \label{sec:conditioning}
Inside music composition with DL there is a task that is the harmonization of a given melody which differs to the task of creating a polyphonic melody from scratch. In one hand, if we analyze the harmony of a created melody from scratch with a DL model, we saw that music generated with DL is not well structured as it does not compose different sections and write aesthetic cadences or bridges between the sections in and end-to-end way yet. In spite of that, the harmony generated by Transformer-based models that compose polyphonic melodies is coherent in the first bars of the generated pieces \cite{music_transformer} because it follows a certain key. We have to emphasize here that these melodies are written for piano, which differs from multi-instrument music that presents added challenges such as generating appropriate melodies or accompaniments for each instrument or  deciding which instruments make up the ensemble (see section \ref{sec:instrumentation}).

In the other hand, the task of melody harmonization consists of generating the harmony that accompanies a given melody. The accompaniment can be a chord accompaniment regardless of the instrument or the track where the chords are, and multi-track accompaniment where the notes in each chord belong to a specific instrument.
Firsts models for harmonization used HMM, but these models were improved by RNNs. Some models predicted chord functions \cite{abs-2001-02360} and other models match chord accompaniments for a given melody \cite{yang2019clstms}. Regarding the generation of accompaniment with different tracks, there have been proposed GAN-based models which implement lead sheet arrangements. In 2018, a Multi-Instrument Co-Arrangement model called MICA \cite{ZhuLYQLZZWXC18} and its improvement MSMICA in 2020 \cite{ZhuLYZZC20} were proposed to generate multi-track accompaniment. There is also a model called the Bach Doodle \cite{HuangHRDWHH19} which used Coconet \cite{counterpoint_conv} to generate accompaniments for a given melody in Bach style. The harmonic quality of these models improves the harmony generated by models that create polyphonic melodies from scratch because the model focus on the melodic content to perform the harmonization which represents a smaller challenge than generating an entire well-harmonized piece from scratch.

There are more tasks inside music generation with DL using conditioning, such as generating a melody given a chord progression, which is a way of composing that humans follow. This tasks have been addressed with Variational AutoEncoders (VAEs) \cite{TengZG17}, Generative Adversarial Networks or GAN-based models \cite{jazzgan}, \cite{bebopnet}\footnote{\url{https://shunithaviv.github.io/bebopnet/}, accessed August 2021} and end-to-end models \cite{ZhuLYQLZZWXC18}. Other models perform the full process of composing, such as ChordAL \cite{Tan19}. This model generates chords and then the obtained chord progression is sent to a melody generator and the final output is sent to a music style processor. Models like BebopNet \cite{bebopnet} generate a melody from jazz chords because this style presents additional challenges in the harmony context.

\subsection{Genre Transformation with Style Transfer} \label{sec:style_transfer}
In music, a style or genre is defined as a complex mixture of features ranging from music theory to sound design. These features include timbre, the composition process and the instruments used in the music piece or the effects with which music is synthesized. Due to the fact that there are lots of music genres and the lack of datasets for some of those genres, it is common to use style transfer techniques to transform music in a determined style into other style by changing the pitch of  existing notes or adding new instruments that fit in the style into which we want to transform the music.

In computer-based music composition, the most common technique for performing style transfer in music is to obtain an embedding of the style and the use this embedding or feature vector to generate new music. Style transfer \cite{style_transfer} in NNs was introduced in 2016 by Gatys et al. with the idea of applying style features to an image from another image. 
One of the first studies that used style transfer for symbolic music generation was MIDI-VAE \cite{midi_vae} in 2018. MIDI-VAE encodes the style in the latent space as a combination of pitch, dynamics and instrument features to generate polyphonic music. Style transfer can also be achieved with Transfer Learning \cite{ZhuangQDXZZXH21}. The first work that used transfer learning to perform style transfer was a recurrent VAE model for jazz which was proposed by Hung et al. \cite{HungWYW19} in 2019. Transfer learning is done by training the model on the source dataset, and then fine-tuning the resulting model parameters on the target dataset that can be in a different style than the source dataset. This model showed that using Transfer Learning to transform a music piece in a determined style to another is a great solution because it could not only be used to transform existing pieces in new genres but it could also be used to compose music from scratch in genres that are not present in the music composition datasets that are being used nowadays. An example of this could be the use a NN trained with a large dataset for pop such as the Lakh MIDI Dataset (LMD) \cite{raffel2016learning} and use this pre-trained model to generate urban music though Transfer Learning.

Other music features such as harmony and texture (see fig. \ref{fig:composing_scheme}) have been also used as style transfer features \cite{abs-2008-07122}, \cite{pianotree_vae}, \cite{muse_morphose}. There have also been studied fusion genres models in which different styles are mixed to generate music in an unknown style \cite{ChenWLLL17}.

\section{Instrumentation and Orchestration} \label{sec:instrumentation}
As we mentioned in section \ref{sec:process}, Instrumentation and orchestration are fundamental elements in the musical genre being composed, and may represent a characteristic signature of each composer by the use of specific instruments or the way their compositions are orchestrated. An example of that is the orchestration that Beethoven used in his symphonies that changed the way in which music was composed \cite{grove1962beethoven}. \textit{Instrumentation} is the study of how to combine similar or different instruments in varying numbers in order to create an ensemble \textit{Orchestration} is the selection and combination of similarly or differently scored sections \cite{sevsay2013cambridge}.
From that, we can relate Instrumentation as the color of a piece and Orchestration to the aesthetic aspect of the composition. Instrumentation and Orchestration have a huge impact on the way we perceive music and so, to the emotional part of music, but, although they represent a fundamental part of the music, emotions are beyond the scope of this work.

\subsection{From Polyphonic to Multi-Instrument Music Generation} \label{sec:scratch}
\begin{figure}
    \centering
    \includegraphics[width=0.85\columnwidth]{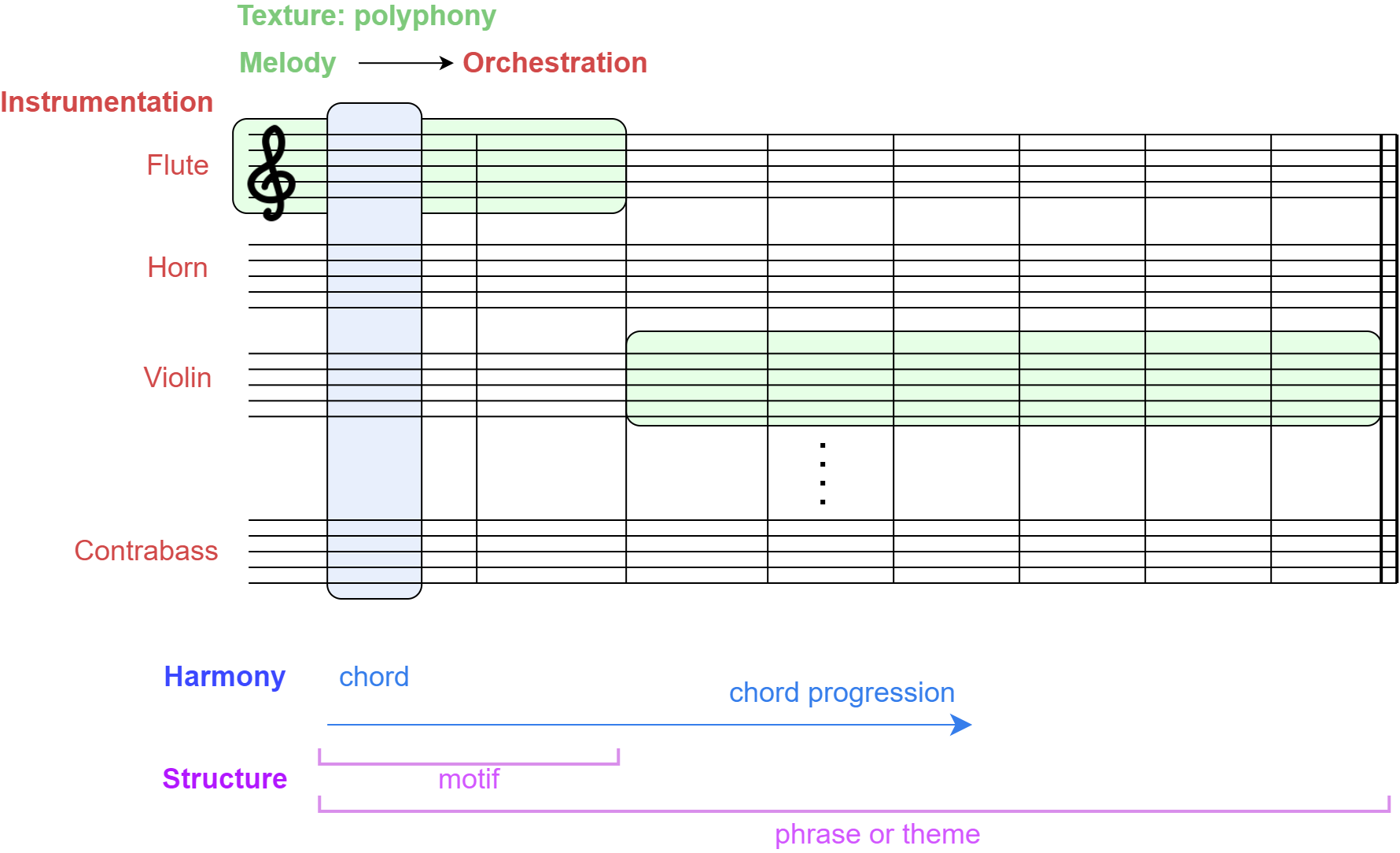}
    \caption{Scheme of an output-like score of multi-instrument generation models.}
    \label{fig:multiinstrument_scheme}
\end{figure}

In computer-based music composition we can group Instrumentation and Orchestration concepts in multi-instrument or multi-track music. However, the DL-based models for multi-instrument generation do not work exactly with those concepts. Multi-instrument DL-based models generate polyphonic music for more than one instrument but, does the generated music follow a coherent harmonic progression? Is the resulting arrangement coherent in terms of Instrumentation and Orchestration or do DL-based models just generate multi-instrument music not taking into account the color of each instrument or the arrangement? In section \ref{sec:melody} we showed that polyphonic music generation could compose music with a certain harmonic sense, but when facing multi-instrument music one of the most important aspects to take into account is the color of the instruments and the ensemble. Deciding how many and which instruments are in the ensemble, and how to divide the melody and accompaniment between them is not yet a solved problem in Music Generation with DL. In the recent years this challenges have been faced by building DL models that generate music from scratch that can be interactive models in which humans can select the instruments of the ensemble \cite{ens2020mmm}. There are also models that allows to inpaint instruments or bars. We describe these models and answer to the exposed questions in section \ref{sec:scratch}. 
In Fig. \ref{fig:multiinstrument_scheme} we show the scheme with the music basic principles of an output-like score of a multi-instrument generation model. 

\subsection{Multi-Instrument Generation from Scratch} \label{sec:scratch}
The first models that could generate multi-track music have been proposed rencently. Previous to multi-track music generation, some models generated the drums track for a given melody or chords. An example of those models are the model proposed in 2012 by Kang et al. \cite{semin2012automatic} which accompanied a melody in a given scale with an automated drums generator. Later on, in 2017 Chu et al. \cite{song_pi} used a hierarchical RNN to generate pop music in which drums were present.

\begin{figure}[h]
    \centering
    \includegraphics[height=5cm, width=0.7\textwidth]{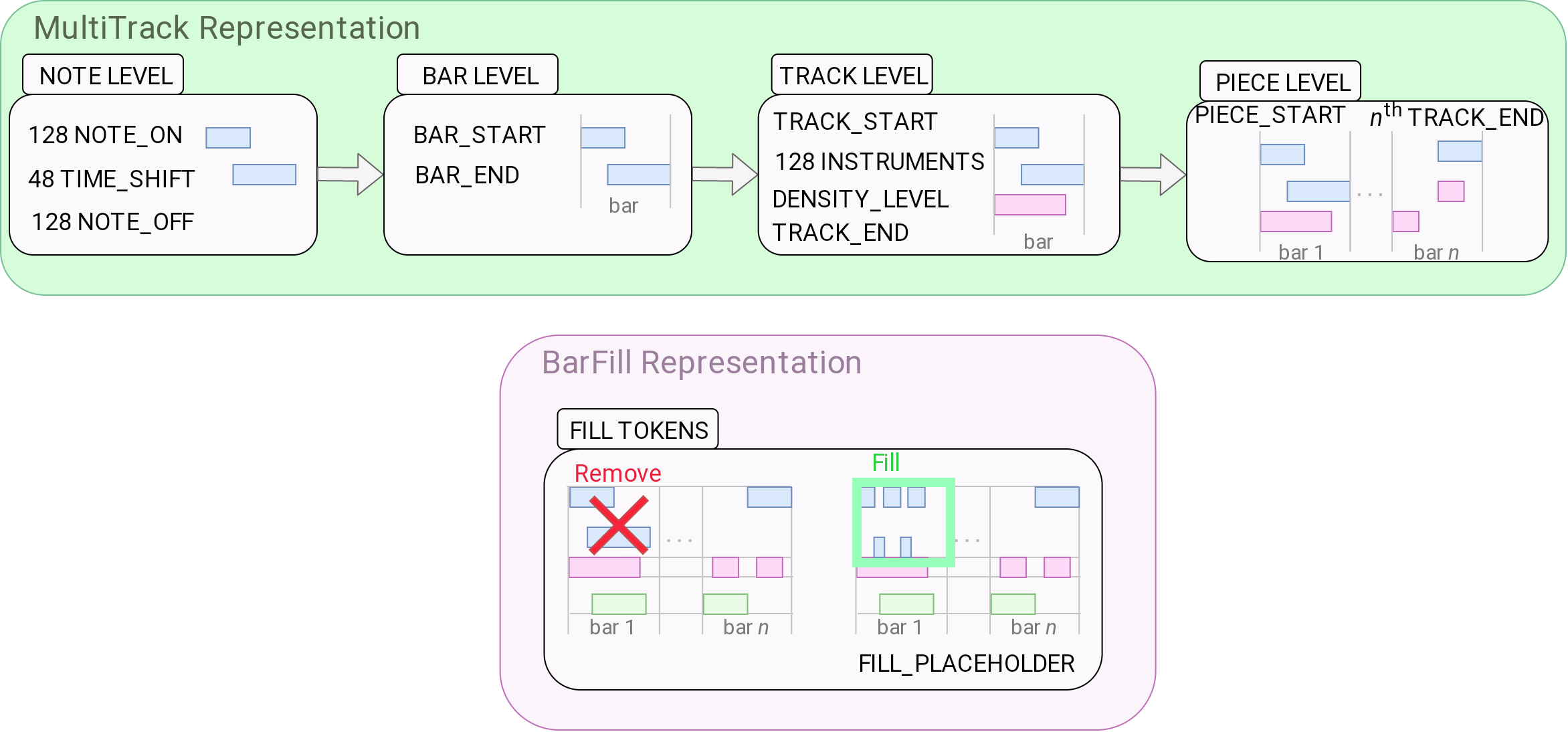}
    \caption{MMM token representations reproduced from \cite{ens2020mmm}}
    \label{fig:mmm}
\end{figure}

One of the most used architectures in music generation are the generative models such as GANs and VAEs. The first considered and most well-known model for multi-track music generation is MuseGAN \cite{musegan}, presented in 2017. Then, more models followed the multi-instrument generation task \cite{seqgan}, \cite{binary} and later in 2020 other models based on AutoEncoders such as MusAE \cite{musae} were released. The other big group of NN architectures that have being used recently to generate music are the Transformers. The most well-known model for music generation with Transformers is the Music Transformer \cite{music_transformer} for piano polyphonic music generation. In 2019, Donahue et al. \cite{lahknes} proposed LakhNES for multi-track music generation and in 2020, Ens et al. proposed a Conditional Multi-Track Music Generation model (MMM) \cite{ens2020mmm} which is based on LakhNES and improves the token representation of this previous model by concatenating multiple tracks into a single sequence. This model uses a MultiInstrument and a BarFill representations which are represented in Fig. \ref{fig:mmm}. In Fig. \ref{fig:mmm} we show the MultiInstrument representation which contains the tokens that the MMM model uses for generating music, and the BarFill representation that is used to inpaint, that is, generating a bar or a couple of bars but maintaining the rest of the composition.

From the composition process point of view, these models do not orchestrate or instrumentate, but they create music from scratch or by inpainting. This means that these models do not choose the number of instruments and do not generate high-quality melodic or accompaniment content related to the instrument that is being selected. As an example, the MMM model generates melodic content for a predefined instrument which follows the timbral features of the instrument, but when inpainting or recreating a single instruments while keeping the other tracks, it is sometimes difficult to follow the key in which the other instruments are composed. This leads us to the conclusion that multi-instrumental models for music generation focus on end-to-end generation, but still do not work well when it comes to instrumentation or orchestration, as they still cannot decide the number of instruments in the generated piece of music. They generate music for the ensemble they were trained on, such as LakhNES \cite{lahknes} or they take predefined tracks to generate the content of each track \cite{ens2020mmm}. More recent models, such as MMM, are opening up interactivity between human and artificial intelligence in terms of multi-instrument generation, which will allow better tracking of the human compositional process and thus improve the music generated with multiple instruments.

\section{Evaluation and Metrics} \label{sec:evaluation}
Evaluation in music generation can be divided according to the way in which the output of a DL mode is measured. Ji et al. \cite{survey} differentiate between the evaluation from the objective point of view and from the subjective point of view. In music, it is necessary to measure the results from a subjective point of view because it is the type of evaluation that tells us how much creativity the model brings compared to human creativity. Sometimes, the objective evaluation that calculates the metrics of the results of a model can give us an idea of the quality of these results, but it is difficult to find a way to relate it to the concept of creativity. In this section we show how the state-of-the-art models measure the quality of their results from objective and subjective points of view.

\subsection{Objective Evaluation}
The objective evaluation measures the performance of the model and the quality of its outputs using some numerical metrics. In music generation there is the problem of comparing models trained for different purposes and models trained with different datasets, so we give a description of the most common metrics used in state-of-the-art models. Ji et al. \cite{survey} differentiate between model metrics and music metrics or descriptive statistics, and other methods such as pattern repetition \cite{WangD14} or plagiarism detection \cite{deepbach}.

When you want to measure the performance of a model, the most used metrics depending on the DL model used to generate music are: the loss, the perplexity, the BLEU score, the precision (P), recall (R) or F-score (F$_1$). Normally, these metrics are used to compare different DL models that are built for the same purpose. 

Loss is usually used to show the difference between the inputs and outputs of the model from a mathematical point of view, while, on the other hand, perplexity tells us the generalization capability that a model has, which is more related to how the model generates new music. As an example, the Music Transformer \cite{music_transformer} use the loss and the perplexity to compare the outputs between different Transformer architectures in order to validate the model, TonicNet \cite{tonicnet} only uses the loss for the same purpose and MusicVAE \cite{musicvae} only uses a measure that indicates the quality of reconstruction that the model has, but does not use any metric to compare between other DL music generation models.

Regarding metrics that are specifically related to music, that is, those that take into account musical descriptors, we can find that these metrics help to measure the quality of a composition. According to Ji et al. \cite{survey} these metrics can be grouped in four categories: pitch-related, rhythm-related, harmony-related and style transfer-related. Pitch-related metrics \cite{survey} such as scale consistency, tone spam, ratio of empty bars or number of pitch classes used, are metrics that measures pitch attributes in general. Rhythm-related metrics take into account the duration or pattern of the notes and are, for example, the rhythm variations, the number of concurrent three or four notes or the duration of repeated pitches. The harmony-related metrics measure the chords entropy, distance or coverage. These three metric categories are used by models like MuseGAN \cite{musegan}, C-RNN-GAN \cite{crnngan} or JazzGAN \cite{jazzgan}. Finally, techniques related to style transfer help to understand how close or far the generation is from the desired style. These include, among others, the style fit, the content preservation or the strength of transfer \cite{BrunnerWWZ18}.

\subsection{Subjective Evaluation}
The subjective point of view determines how the generated music is in terms of creativity and novelty, that is, to what extent the music generated can be considered art. There is no a way to define art, although art involves creativity and aesthetics. Sternberg and Kaufman \cite{sternberg2018nature} defined creativity as the ability to make contributions that are both novel and appropriate to the task, often with an added component such as being quality, surprising, or useful. Creativity requires a deeper understanding of the nature and use of musical knowledge. According to Ji et al. \cite{survey} there is a lack of correlation between the quantitative evaluation of music quality and human judgement, which means that music generation models must be also evaluated from a subjective point of view which would give us insights of the creativity of the model. The most used method in subjective evaluations is the listening test which often consists on humans trying to differentiate between machine-generated or human-created music. This method is known as the Turing test which is performed to test DeepBach \cite{deepbach}. In this model, 1.272 people belonging to different musical experience groups took the test. This test showed that the more complex the model was, the better outputs it got. MusicVAE \cite{musicvae} also perform a listening test and the Kruskal Wallis H-Test to validate the quality of the model, which conclude that the model performed better with the hierarchical decoder. 
MuseGAN \cite{musegan} also conducted a listening test with 144 users divided into groups with different musical experience, but there were pre-defined questions that the users had to vote in a range from 1 to 5: the harmony complaisance, the rhythm, the structure, the coherence and the overall rating.

Other listening methods require scoring the generated music, this is known as side-by-side rating \cite{survey}. It is also possible to ask some questions to the listeners about the creativity of the model or the naturalness of the generated piece, among other questions, depending on the generation goal of the model. One important thing to keep in mind in listening tests is the variability of the population that is chosen for the test (if listeners are music students with a basic knowledge of music theory, if they are amateurs and so they do not have any music knowledge or if they are professional musicians). The listeners must have the same stimuli and also listen to the same pieces and have as reference (if it applicable) the same human-created pieces. Auditory fatigue must also be taken into account as there can be an induced bias in the listeners if the listen to similar samples for a long period of time. 

Being said that, we can conclude that listening tests are indispensable when it comes to music generation because it gives a feedback of the quality of the model and they can also be a way to find the better NN architecture or DL model that is being studied.

\section{Discussion} \label{sec:discussion}

We have showed that music is a structured language model with temporal and harmonic short and long-term relationships which requires a deep understanding of all of its insights in order to be modelled. This, in addition to the variety of genres and subgenres that exist in music and the large number of composing strategies that can be followed to compose a music piece, makes the field of music generation with DL a constantly growing and challenging field. Having described the music composition process and recent work in DL for music generation, we will now address the issues raised in the section \ref{sec:challenges}.

\subsubsection*{Are the current DL models capable of generating music with a certain level of creativity?}
The first models that generated music with DL used RNNs such as LSTMs. These models could generate notes but they failed when they generated long-term sequences. This was due to the fact that these NNs did not handle the long-term sequences that are required for music generation. In order to solve this problem and being able to generate short motifs by interpolating two existing motifs or sampling from a distribution, MusicVAE was created. But some questions arise from here: Do interpolations between existing motifs generate high quality ones which have sense inside the same music piece? If we use MusicVAE for creating a short motif we could get very good results, but if we use this kind of models to generate longer phrases or motifs that are similar to the inputs, these interpolations may output motifs that can be aesthetic but sometimes they do not follow any rhythmic or notes direction (ascendent or descendent) pattern that the inputs have. Therefore, these interpolations usually cannot generate high quality motifs because the model does not understand the rhythmic patterns and notes directions. In addition, chord progressions normally do have inversions and there are rules in classical music or stylistic restrictions in pop, jazz or urban music that determine how each chord is followed by another chord. If we analyze the generated polyphonic melodies of DL methods there is a lack of quality in terms of harmonic content, because NNs which are trained to generate music could not understand all these intricacies that are present in the music language or because this information should be passed to the NN as part of the input for example as tokens.

\subsubsection*{What is the best NN architecture to perform music composition with DL?} The Transformer architecture has been used with different attention mechanisms that allow longer sequences to be modeled. An example of this is the success of the MMM model \cite{ens2020mmm} that used GPT-2 to generate multi-track music. Despite the fact that the model used a pre-trained Transformer for text generation, it generates coherent music in terms of harmony and rhythm.
Other architectures use Generative Networks such as GANs or VAEs, and also a combination of these architectures with Transformers. The power that bring these models is the possibility to extract high-level music attributes such as the style, and low-level features that are organized in a latent space. This latent space is then used to interpolate between those features and attributes to generate new music based on existing pieces and music styles.

Analyzing the NN models and architectures that have been used in the past years to generate music with DL, there is not a specific NN architecture that performs better for this purpose because the best NN architecture that can be used to build a music generation model will depend on the output that we want to obtain. In spite of that, Transformers and Generative models are are emerging as the best alternatives in this moment as the latest works in this domain demonstrates. A combination of both models is also a great option to perform music generation \cite{transformer_vae} although it depends on the output we want to generate, and sometimes the best solution comes from a combination of DL with probabilistic methods.
Another aspect to take into account is that generally, music generation requires models with a large number of parameters and data. We can solve this problem by taking a pre-trained model as some state-of-the-art models which we have described in the previous sections, and then perform fine-tuning to another NN architecture. Another option is having a pre-trained latent space which has been generated by training a big model with a huge dataset like MusicVAE proposes, and then train a smaller NN with less data taking advantage of the pre-trained latent space in order to condition the style of the music compositions as MidiMe proposes \cite{midime}.

\subsubsection*{Could end-to-end methods generate entire structured music pieces?}
As we described in section \ref{sec:structure}, nowadays there are structure template based models that can generate structured music \cite{Lattner_thesis}, but there is not yet an end-to-end method that can compose a structured music piece. The music composition process that a human composer follows is similar to this template-based method. In the near future, it would be likely that AI could compose structured music from scratch, but the question here is whether AI models for music generation will be used to compose entire music pieces from scratch or whether these models would be more useful as an aid to composers and thus as an interaction between humans and AI.

\subsubsection*{Are the composed pieces with DL just an imitation of the inputs or can NNs generate new music in styles that are not present in the training data?}
When training DL models, some information that is in the input which is passed to the NN can be present without any modification in the output. Even that, MusicVAE \cite{musicvae} and other DL models for music generation showed that new music can be composed without imitating existing music or committing plagiarism. Imitate the inputs could be a case of overfitting which is never the goal of a DL model. It should also be taken into account that it is very difficult to commit plagiarism in the generation of music due to the great variety of instruments, tones, rhythms or chords that may be present in a piece of music.

\subsubsection*{Should NNs compose music by following the same logic and process as humans do?}
We showed that researchers started to build models that could generate polyphonic melodies but these melodies did not follow any direction after a few bars. When MusicVAE came out, it was possible to generate high quality motifs and this encouraged new studies to generate melodies taking information of past time steps. New models such as Diffusion Models \cite{diffusion_models} are using this pre-trained models to generate longer sequences to let melodies follow patterns or directions. We also show that there are models that can generate a melody by conditioning it with chord progressions which is the way of composing music in styles like pop. Comparing the human way of composing to the DL architectures that are being used to generate music, we can see some similarities of both processes specially in auto-regressive models. Auto-regression (AR) consists on predict the future values from past events. Some DL methods are auto-regressive, and the fact that new models are trying to generate longer sequences by taking information of past time steps resemble the human composing process of classical music.

\subsubsection*{How much data do DL models for music generation need?} This question can be answered partially if we look at the state-of-the-art models. MusicVAE uses the LMD \cite{raffel2016learning} with 3.7 million melodies, 4.6 million drum patterns and 116 thousand trios. Music Transformer instead uses only 1.100 piano pieces from the Piano-e-Competition to train the model. Other models such as the MMM takes the GPT-2 which is a pre-trained Transformer with lots of text data. This lead us to affirm that DL models for music generation do need lots of data, specially when trainig Generative Models or Transformers, but taking pre-trained models and perform transfer learning is also a good solution specially for music genres and subgenres that are not represented in the actual datasets for symbolic music generation. 

\subsubsection*{Are current evaluation methods good enough to compare and measure the creativity of the composed music?} As we described in section \ref{sec:evaluation}, there are two evaluation categories: the objective and the subjective evaluation. Objective evaluation metrics are similar between existing methods but there is a lack of a general subjective evaluation method. The listening tests are the most used subjective evaluation method but sometimes the Turing test which just asks to distinguish between a computer-based or a human composition is not enough to know all the characteristics of the compositions created by a NN. The solution to this problem would be to ask general questions related to the quality of the music features showed in fig. \ref{fig:composing_scheme} as MuseGAN proposes, and use the same questions and the same rating method in the DL models to set a general subjective evaluation method.

\section{Conclusions and Future Work} \label{sec:future_work}

In this paper, we have described the state-of-the-art in DL music generation by giving an overview of the NN architectures that have been used for DL music generation, and discussed the challenges that are still open in the use of deep NNs in music generation.

The use of DL architectures and techniques for the generation of music (as well as other artistic content) is a growing area of research. However, there are open challenges such as generating music with structure, analyzing the creativity of the generated music and build interactivity models that could help composers. Future work should focus on better modeling the long-term relationships (in time and harmony axes) in order to generate well-structured and harmonized music that does not get loose after a few bars, and inpainting or human-AI interaction which is a task with a growing interest in the recent years. There is also a pending challenge that has to do with transfer learning or the conditioning of the generation of styles that allows not to be restricted only to the same authors and genres that are present in the publicly available datasets, such as the JSB Chorales Dataset or the Lakh MIDI Dataset, which makes most of the state-of-the-art works only focus on the same music styles. When it comes to multi-instrument generation, this task does not follow the human composing process and it could be interesting to see new DL models that first compose a high-quality melodic content and then decide by themselves or with human's help the number of instruments of the music piece and be able to write high-quality music for each instrument attending to its timbral features.  Further questions related to the directions that music generation with DL should focus on, that is, building end-to-end models that can generate high-creative music from scratch or interactive models in which composers could interact with the AI is a task that future will solve, although the trending of human-AI interaction is growing faster everyday.

There are more open questions in music composition with DL that are not in the scope of this paper. Questions like who owns the intellectual property of music generated with DL if the NNs are trained with copyrighted music. We suggest that this would be an important key in commercial applications. The main key here is to define what makes a composition different from others and there are several features that play an important role here. As we mentioned in the section \ref{sec:introduction}, these features include the composition itself, but also the timbre and effects used in creating the sound of the instruments. 
From the point of view of composition, which is the scope of our study, we can state that when generating music with DL it is always likely to generate music that is similar to the inputs, and sometimes the music generated has patterns taken directly from the inputs, so further research would have to be done in this area from the point of view of music theory, intellectual property and science to define what makes a composition different from others and how music generated with DL could be registered.

We hope that the analysis presented in this article will help to better understand the problems and possible solutions and thus may contribute to the overall research agenda of deep learning-based music generation.

\subsubsection*{Acknowledgments.}
This research has been partially supported by the Spanish Science, Innovation and University Ministry by the RTI2018-096986-B-C31 contract and the Aragonese Government by the AffectiveLab-T60-20R project.

We wish to thank J\"urgen Schmidhuber for his suggestions.

\bibliographystyle{unsrt}  

\bibliography{references}

\begin{thebibliography}{10}

\bibitem{definition}
\url{https://www.copyright.gov/prereg/music.html}, 2019.
\newblock accessed July 2021.

\bibitem{chomsky2009syntactic}
Noam Chomsky.
\newblock {\em Syntactic structures}.
\newblock De Gruyter Mouton, 2009.

\bibitem{cope_emi}
David Cope.
\newblock Experiments in musical intelligence (emi): Non-linear
  linguistic-based composition.
\newblock {\em Journal of New Music Research}, 18(1-2):117--139, 1989.

\bibitem{xenakis1981musiques}
Iannis Xenakis.
\newblock Musiques formelles nouveaux principes formels de composition
  musicale.
\newblock 1981.

\bibitem{Koening}
\url{https://koenigproject.nl/project-1/}, 2019.
\newblock accessed July 2021.

\bibitem{Todd1989}
Peter~M. Todd.
\newblock A connectionist approach to algorithmic composition.
\newblock {\em Computer Music Journal}, 13(4):27--43, 1989.

\bibitem{nierhaus2009algorithmic}
Gerhard Nierhaus.
\newblock {\em Algorithmic composition: paradigms of automated music
  generation}.
\newblock Springer Science \& Business Media, 2009.

\bibitem{hiller1957musical}
Lejaren~A Hiller~Jr and Leonard~M Isaacson.
\newblock Musical composition with a high speed digital computer.
\newblock In {\em Audio Engineering Society Convention 9}. Audio Engineering
  Society, 1957.

\bibitem{deepbach}
Ga{\"{e}}tan Hadjeres, Fran{\c{c}}ois Pachet, and Frank Nielsen.
\newblock Deepbach: a steerable model for bach chorales generation.
\newblock In Doina Precup and Yee~Whye Teh, editors, {\em Proceedings of the
  34th International Conference on Machine Learning, {ICML} 2017, Sydney, NSW,
  Australia, 6-11 August 2017}, volume~70 of {\em Proceedings of Machine
  Learning Research}, pages 1362--1371. {PMLR}, 2017.

\bibitem{Mozer94}
Michael~C. Mozer.
\newblock Neural network music composition by prediction: Exploring the
  benefits of psychoacoustic constraints and multi-scale processing.
\newblock {\em Connect. Sci.}, 6(2-3):247--280, 1994.

\bibitem{Eck01}
Douglas Eck.
\newblock A network of relaxation oscillators that finds downbeats in rhythms.
\newblock In Georg Dorffner, Horst Bischof, and Kurt Hornik, editors, {\em
  Artificial Neural Networks - {ICANN} 2001, International Conference Vienna,
  Austria, August 21-25, 2001 Proceedings}, volume 2130 of {\em Lecture Notes
  in Computer Science}, pages 1239--1247. Springer, 2001.

\bibitem{EckS02}
Douglas Eck and J{\"{u}}rgen Schmidhuber.
\newblock Learning the long-term structure of the blues.
\newblock In Jos{\'{e}}~R. Dorronsoro, editor, {\em Artificial Neural Networks
  - {ICANN} 2002, International Conference, Madrid, Spain, August 28-30, 2002,
  Proceedings}, volume 2415 of {\em Lecture Notes in Computer Science}, pages
  284--289. Springer, 2002.

\bibitem{bharucha1989modeling}
Jamshed~J Bharucha and Peter~M Todd.
\newblock Modeling the perception of tonal structure with neural nets.
\newblock {\em Computer Music Journal}, 13(4):44--53, 1989.

\bibitem{ZhuangQDXZZXH21}
Fuzhen Zhuang, Zhiyuan Qi, Keyu Duan, Dongbo Xi, Yongchun Zhu, Hengshu Zhu, Hui
  Xiong, and Qing He.
\newblock A comprehensive survey on transfer learning.
\newblock {\em Proc. {IEEE}}, 109(1):43--76, 2021.

\bibitem{vae}
Diederik~P. Kingma and Max Welling.
\newblock Auto-encoding variational bayes.
\newblock In Yoshua Bengio and Yann LeCun, editors, {\em 2nd International
  Conference on Learning Representations, {ICLR} 2014, Banff, AB, Canada, April
  14-16, 2014, Conference Track Proceedings}, 2014.

\bibitem{beta_vae}
Irina Higgins, Lo{\"{\i}}c Matthey, Arka Pal, Christopher Burgess, Xavier
  Glorot, Matthew Botvinick, Shakir Mohamed, and Alexander Lerchner.
\newblock beta-vae: Learning basic visual concepts with a constrained
  variational framework.
\newblock In {\em 5th International Conference on Learning Representations,
  {ICLR} 2017, Toulon, France, April 24-26, 2017, Conference Track
  Proceedings}. OpenReview.net, 2017.

\bibitem{musicvae}
Adam Roberts, Jesse~H. Engel, Colin Raffel, Curtis Hawthorne, and Douglas Eck.
\newblock A hierarchical latent vector model for learning long-term structure
  in music.
\newblock In Jennifer~G. Dy and Andreas Krause, editors, {\em Proceedings of
  the 35th International Conference on Machine Learning, {ICML} 2018,
  Stockholmsm{\"{a}}ssan, Stockholm, Sweden, July 10-15, 2018}, volume~80 of
  {\em Proceedings of Machine Learning Research}, pages 4361--4370. {PMLR},
  2018.

\bibitem{gan}
Ian~J. Goodfellow, Jean Pouget-Abadie, Mehdi Mirza, Bing Xu, David
  Warde-Farley, Sherjil Ozair, Aaron Courville, and Yoshua Bengio.
\newblock Generative adversarial nets.
\newblock NIPS'14, page 2672–2680, Cambridge, MA, USA, 2014. MIT Press.

\bibitem{rosenblatt1958perceptron}
Frank Rosenblatt.
\newblock The perceptron: a probabilistic model for information storage and
  organization in the brain.
\newblock {\em Psychological review}, 65(6):386, 1958.

\bibitem{lstm}
Sepp Hochreiter and J{\"u}rgen Schmidhuber.
\newblock Long short-term memory.
\newblock {\em Neural computation}, 9(8):1735--1780, 1997.

\bibitem{FukushimaM82}
Kunihiko Fukushima and Sei Miyake.
\newblock Neocognitron: {A} new algorithm for pattern recognition tolerant of
  deformations and shifts in position.
\newblock {\em Pattern Recognit.}, 15(6):455--469, 1982.

\bibitem{CunHBB99}
Yann LeCun, Patrick Haffner, L{\'{e}}on Bottou, and Yoshua Bengio.
\newblock Object recognition with gradient-based learning.
\newblock In David~A. Forsyth, Joseph~L. Mundy, Vito~Di Ges{\`{u}}, and Roberto
  Cipolla, editors, {\em Shape, Contour and Grouping in Computer Vision},
  volume 1681 of {\em Lecture Notes in Computer Science}, page 319. Springer,
  1999.

\bibitem{attention}
Ashish Vaswani, Noam Shazeer, Niki Parmar, Jakob Uszkoreit, Llion Jones,
  Aidan~N. Gomez, Lukasz Kaiser, and Illia Polosukhin.
\newblock Attention is all you need.
\newblock In Isabelle Guyon, Ulrike von Luxburg, Samy Bengio, Hanna~M. Wallach,
  Rob Fergus, S.~V.~N. Vishwanathan, and Roman Garnett, editors, {\em Advances
  in Neural Information Processing Systems 30: Annual Conference on Neural
  Information Processing Systems 2017, December 4-9, 2017, Long Beach, CA,
  {USA}}, pages 5998--6008, 2017.

\bibitem{music_transformer}
Cheng-Zhi~Anna Huang, Ashish Vaswani, Jakob Uszkoreit, Noam Shazeer, Curtis
  Hawthorne, Andrew~M Dai, Matthew~D Hoffman, and Douglas Eck.
\newblock Music transformer: Generating music with long-term structure.
\newblock {\em arXiv preprint arXiv:1809.04281}, 2018.

\bibitem{levi1991field}
Raymond~Glenn Levi.
\newblock {\em A field investigation of the composing processes used by
  second-grade children creating original language and music pieces}.
\newblock PhD thesis, Case Western Reserve University, 1991.

\bibitem{collins2005synthesis}
David Collins.
\newblock A synthesis process model of creative thinking in music composition.
\newblock {\em Psychology of music}, 33(2):193--216, 2005.

\bibitem{walton2005basic}
Charles~W Walton.
\newblock {\em Basic Forms in Music}.
\newblock Alfred Music, 2005.

\bibitem{ens2020mmm}
Jeff Ens and Philippe Pasquier.
\newblock Mmm: Exploring conditional multi-track music generation with the
  transformer.
\newblock {\em arXiv preprint arXiv:2008.06048}, 2020.

\bibitem{nades}
Hugo Larochelle and Iain Murray.
\newblock The neural autoregressive distribution estimator.
\newblock In Geoffrey~J. Gordon, David~B. Dunson, and Miroslav Dud{\'{\i}}k,
  editors, {\em Proceedings of the Fourteenth International Conference on
  Artificial Intelligence and Statistics, {AISTATS} 2011, Fort Lauderdale, USA,
  April 11-13, 2011}, volume~15 of {\em {JMLR} Proceedings}, pages 29--37.
  JMLR.org, 2011.

\bibitem{unitselection}
Mason Bretan, Gil Weinberg, and Larry~P. Heck.
\newblock A unit selection methodology for music generation using deep neural
  networks.
\newblock In Ashok~K. Goel, Anna Jordanous, and Alison Pease, editors, {\em
  Proceedings of the Eighth International Conference on Computational
  Creativity, {ICCC} 2017, Atlanta, Georgia, USA, June 19-23, 2017}, pages
  72--79. Association for Computational Creativity {(ACC)}, 2017.

\bibitem{waite2016generating}
Elliot Waite et~al.
\newblock Generating long-term structure in songs and stories.
\newblock {\em Web blog post. Magenta}, 15(4), 2016.

\bibitem{abs-1709-06404}
Ga{\"{e}}tan Hadjeres and Frank Nielsen.
\newblock Interactive music generation with positional constraints using
  anticipation-rnns.
\newblock {\em CoRR}, abs/1709.06404, 2017.

\bibitem{raffel2016learning}
Colin Raffel.
\newblock {\em Learning-based methods for comparing sequences, with
  applications to audio-to-midi alignment and matching}.
\newblock PhD thesis, Columbia University, 2016.

\bibitem{musenet}
Christine Payne.
\newblock Musenet, 2019.
\newblock {\em URL https://openai. com/blog/musenet}, 2019.

\bibitem{transformer_vae}
Junyan Jiang, Gus Xia, Dave~B. Carlton, Chris~N. Anderson, and Ryan~H.
  Miyakawa.
\newblock Transformer {VAE:} {A} hierarchical model for structure-aware and
  interpretable music representation learning.
\newblock In {\em 2020 {IEEE} International Conference on Acoustics, Speech and
  Signal Processing, {ICASSP} 2020, Barcelona, Spain, May 4-8, 2020}, pages
  516--520. {IEEE}, 2020.

\bibitem{pianotree_vae}
Ziyu Wang, Yiyi Zhang, Yixiao Zhang, Junyan Jiang, Ruihan Yang, Junbo Zhao, and
  Gus Xia.
\newblock {PIANOTREE} {VAE:} structured representation learning for polyphonic
  music.
\newblock {\em CoRR}, abs/2008.07118, 2020.

\bibitem{diffusion_models}
Gautam Mittal, Jesse~H. Engel, Curtis Hawthorne, and Ian Simon.
\newblock Symbolic music generation with diffusion models.
\newblock {\em CoRR}, abs/2103.16091, 2021.

\bibitem{ddpm}
Jonathan Ho, Ajay Jain, and Pieter Abbeel.
\newblock Denoising diffusion probabilistic models.
\newblock In Hugo Larochelle, Marc'Aurelio Ranzato, Raia Hadsell,
  Maria{-}Florina Balcan, and Hsuan{-}Tien Lin, editors, {\em Advances in
  Neural Information Processing Systems 33: Annual Conference on Neural
  Information Processing Systems 2020, NeurIPS 2020, December 6-12, 2020,
  virtual}, 2020.

\bibitem{Lattner_thesis}
Stefan Lattner, Maarten Grachten, and Gerhard Widmer.
\newblock Imposing higher-level structure in polyphonic music generation using
  convolutional restricted boltzmann machines and constraints.
\newblock {\em CoRR}, abs/1612.04742, 2016.

\bibitem{muller2015fundamentals}
Meinard M{\"{u}}ller.
\newblock {\em Fundamentals of Music Processing - Audio, Analysis, Algorithms,
  Applications}.
\newblock Springer, 2015.

\bibitem{chen2019effect}
Ke~Chen, Weilin Zhang, Shlomo Dubnov, Gus Xia, and Wei Li.
\newblock The effect of explicit structure encoding of deep neural networks for
  symbolic music generation.
\newblock In {\em 2019 International Workshop on Multilayer Music
  Representation and Processing (MMRP)}, pages 77--84. IEEE, 2019.

\bibitem{abs-2001-02360}
Yin{-}Cheng Yeh, Wen{-}Yi Hsiao, Satoru Fukayama, Tetsuro Kitahara, Benjamin
  Genchel, Hao{-}Min Liu, Hao{-}Wen Dong, Yian Chen, Terence Leong, and
  Yi{-}Hsuan Yang.
\newblock Automatic melody harmonization with triad chords: {A} comparative
  study.
\newblock {\em CoRR}, abs/2001.02360, 2020.

\bibitem{yang2019clstms}
Wei Yang, Ping Sun, Yi~Zhang, and Ying Zhang.
\newblock Clstms: A combination of two lstm models to generate chords
  accompaniment for symbolic melody.
\newblock In {\em 2019 International Conference on High Performance Big Data
  and Intelligent Systems (HPBD\&IS)}, pages 176--180. IEEE, 2019.

\bibitem{ZhuLYQLZZWXC18}
Hongyuan Zhu, Qi~Liu, Nicholas~Jing Yuan, Chuan Qin, Jiawei Li, Kun Zhang,
  Guang Zhou, Furu Wei, Yuanchun Xu, and Enhong Chen.
\newblock Xiaoice band: {A} melody and arrangement generation framework for pop
  music.
\newblock In Yike Guo and Faisal Farooq, editors, {\em Proceedings of the 24th
  {ACM} {SIGKDD} International Conference on Knowledge Discovery {\&} Data
  Mining, {KDD} 2018, London, UK, August 19-23, 2018}, pages 2837--2846. {ACM},
  2018.

\bibitem{ZhuLYZZC20}
Hongyuan Zhu, Qi~Liu, Nicholas~Jing Yuan, Kun Zhang, Guang Zhou, and Enhong
  Chen.
\newblock Pop music generation: From melody to multi-style arrangement.
\newblock {\em {ACM} Trans. Knowl. Discov. Data}, 14(5):54:1--54:31, 2020.

\bibitem{HuangHRDWHH19}
Cheng{-}Zhi~Anna Huang, Curtis Hawthorne, Adam Roberts, Monica Dinculescu,
  James Wexler, Leon Hong, and Jacob Howcroft.
\newblock Approachable music composition with machine learning at scale.
\newblock In Arthur Flexer, Geoffroy Peeters, Juli{\'{a}}n Urbano, and Anja
  Volk, editors, {\em Proceedings of the 20th International Society for Music
  Information Retrieval Conference, {ISMIR} 2019, Delft, The Netherlands,
  November 4-8, 2019}, pages 793--800, 2019.

\bibitem{counterpoint_conv}
Cheng{-}Zhi~Anna Huang, Tim Cooijmans, Adam Roberts, Aaron~C. Courville, and
  Douglas Eck.
\newblock Counterpoint by convolution.
\newblock In Sally~Jo Cunningham, Zhiyao Duan, Xiao Hu, and Douglas Turnbull,
  editors, {\em Proceedings of the 18th International Society for Music
  Information Retrieval Conference, {ISMIR} 2017, Suzhou, China, October 23-27,
  2017}, pages 211--218, 2017.

\bibitem{TengZG17}
Yifei Teng, Anny Zhao, and Camille Goudeseune.
\newblock Generating nontrivial melodies for music as a service.
\newblock In Sally~Jo Cunningham, Zhiyao Duan, Xiao Hu, and Douglas Turnbull,
  editors, {\em Proceedings of the 18th International Society for Music
  Information Retrieval Conference, {ISMIR} 2017, Suzhou, China, October 23-27,
  2017}, pages 657--663, 2017.

\bibitem{jazzgan}
Nicholas Trieu and R~Keller.
\newblock Jazzgan: Improvising with generative adversarial networks.
\newblock In {\em MUME workshop}, 2018.

\bibitem{bebopnet}
Shunit~Haviv Hakimi, Nadav Bhonker, and Ran El-Yaniv.
\newblock Bebopnet: Deep neural models for personalized jazz improvisations.
\newblock In {\em Proceedings of the 21st international society for music
  information retrieval conference, ismir}, 2020.

\bibitem{Tan19}
Hao~Hao Tan.
\newblock Chordal: {A} chord-based approach for music generation using
  bi-lstms.
\newblock In Kazjon Grace, Michael Cook, Dan Ventura, and Mary~Lou Maher,
  editors, {\em Proceedings of the Tenth International Conference on
  Computational Creativity, {ICCC} 2019, Charlotte, North Carolina, USA, June
  17-21, 2019}, pages 364--365. Association for Computational Creativity
  {(ACC)}, 2019.

\bibitem{style_transfer}
Leon~A. Gatys, Alexander~S. Ecker, and Matthias Bethge.
\newblock Image style transfer using convolutional neural networks.
\newblock In {\em 2016 {IEEE} Conference on Computer Vision and Pattern
  Recognition, {CVPR} 2016, Las Vegas, NV, USA, June 27-30, 2016}, pages
  2414--2423. {IEEE} Computer Society, 2016.

\bibitem{midi_vae}
Gino Brunner, Andres Konrad, Yuyi Wang, and Roger Wattenhofer.
\newblock {MIDI-VAE:} modeling dynamics and instrumentation of music with
  applications to style transfer.
\newblock In Emilia G{\'{o}}mez, Xiao Hu, Eric Humphrey, and Emmanouil Benetos,
  editors, {\em Proceedings of the 19th International Society for Music
  Information Retrieval Conference, {ISMIR} 2018, Paris, France, September
  23-27, 2018}, pages 747--754, 2018.

\bibitem{HungWYW19}
Hsiao{-}Tzu Hung, Chung{-}Yang Wang, Yi{-}Hsuan Yang, and Hsin{-}Min Wang.
\newblock Improving automatic jazz melody generation by transfer learning
  techniques.
\newblock In {\em 2019 Asia-Pacific Signal and Information Processing
  Association Annual Summit and Conference, {APSIPA} {ASC} 2019, Lanzhou,
  China, November 18-21, 2019}, pages 339--346. {IEEE}, 2019.

\bibitem{abs-2008-07122}
Ziyu Wang, Dingsu Wang, Yixiao Zhang, and Gus Xia.
\newblock Learning interpretable representation for controllable polyphonic
  music generation.
\newblock {\em CoRR}, abs/2008.07122, 2020.

\bibitem{muse_morphose}
Shih{-}Lun Wu and Yi{-}Hsuan Yang.
\newblock Musemorphose: Full-song and fine-grained music style transfer with
  just one transformer {VAE}.
\newblock {\em CoRR}, abs/2105.04090, 2021.

\bibitem{ChenWLLL17}
Zhiqian Chen, Chih{-}Wei Wu, Yen{-}Cheng Lu, Alexander Lerch, and Chang{-}Tien
  Lu.
\newblock Learning to fuse music genres with generative adversarial dual
  learning.
\newblock In Vijay Raghavan, Srinivas Aluru, George Karypis, Lucio Miele, and
  Xindong Wu, editors, {\em 2017 {IEEE} International Conference on Data
  Mining, {ICDM} 2017, New Orleans, LA, USA, November 18-21, 2017}, pages
  817--822. {IEEE} Computer Society, 2017.

\bibitem{grove1962beethoven}
George Grove.
\newblock {\em Beethoven and his nine symphonies}, volume 334.
\newblock Courier Corporation, 1962.

\bibitem{sevsay2013cambridge}
Ertu{\u{g}}rul Sevsay.
\newblock {\em The cambridge guide to orchestration}.
\newblock Cambridge University Press, 2013.

\bibitem{semin2012automatic}
Soo-Yol~Ok Semin~Kang and Young-Min Kang.
\newblock Automatic music generation and machine learning based evaluation.
\newblock In {\em International Conference on Multimedia and Signal
  Processing}, pages 436--443, 2012.

\bibitem{song_pi}
Hang Chu, Raquel Urtasun, and Sanja Fidler.
\newblock Song from {PI:} {A} musically plausible network for pop music
  generation.
\newblock In {\em 5th International Conference on Learning Representations,
  {ICLR} 2017, Toulon, France, April 24-26, 2017, Workshop Track Proceedings}.
  OpenReview.net, 2017.

\bibitem{musegan}
Hao{-}Wen Dong, Wen{-}Yi Hsiao, Li{-}Chia Yang, and Yi{-}Hsuan Yang.
\newblock Musegan: Multi-track sequential generative adversarial networks for
  symbolic music generation and accompaniment.
\newblock In Sheila~A. McIlraith and Kilian~Q. Weinberger, editors, {\em
  Proceedings of the Thirty-Second {AAAI} Conference on Artificial
  Intelligence, (AAAI-18), the 30th innovative Applications of Artificial
  Intelligence (IAAI-18), and the 8th {AAAI} Symposium on Educational Advances
  in Artificial Intelligence (EAAI-18), New Orleans, Louisiana, USA, February
  2-7, 2018}, pages 34--41. {AAAI} Press, 2018.

\bibitem{seqgan}
Lantao Yu, Weinan Zhang, Jun Wang, and Yong Yu.
\newblock Seqgan: Sequence generative adversarial nets with policy gradient.
\newblock In Satinder~P. Singh and Shaul Markovitch, editors, {\em Proceedings
  of the Thirty-First {AAAI} Conference on Artificial Intelligence, February
  4-9, 2017, San Francisco, California, {USA}}, pages 2852--2858. {AAAI} Press,
  2017.

\bibitem{binary}
Hao{-}Wen Dong and Yi{-}Hsuan Yang.
\newblock Convolutional generative adversarial networks with binary neurons for
  polyphonic music generation.
\newblock In Emilia G{\'{o}}mez, Xiao Hu, Eric Humphrey, and Emmanouil Benetos,
  editors, {\em Proceedings of the 19th International Society for Music
  Information Retrieval Conference, {ISMIR} 2018, Paris, France, September
  23-27, 2018}, pages 190--196, 2018.

\bibitem{musae}
Andrea Valenti, Antonio Carta, and Davide Bacciu.
\newblock Learning a latent space of style-aware symbolic music representations
  by adversarial autoencoders.
\newblock {\em CoRR}, abs/2001.05494, 2020.

\bibitem{lahknes}
Chris Donahue, Huanru~Henry Mao, Yiting~Ethan Li, Garrison~W. Cottrell, and
  Julian~J. McAuley.
\newblock Lakhnes: Improving multi-instrumental music generation with
  cross-domain pre-training.
\newblock In Arthur Flexer, Geoffroy Peeters, Juli{\'{a}}n Urbano, and Anja
  Volk, editors, {\em Proceedings of the 20th International Society for Music
  Information Retrieval Conference, {ISMIR} 2019, Delft, The Netherlands,
  November 4-8, 2019}, pages 685--692, 2019.

\bibitem{survey}
Shulei Ji, Jing Luo, and Xinyu Yang.
\newblock A comprehensive survey on deep music generation: Multi-level
  representations, algorithms, evaluations, and future directions.
\newblock {\em CoRR}, abs/2011.06801, 2020.

\bibitem{WangD14}
Cheng{-}i Wang and Shlomo Dubnov.
\newblock Guided music synthesis with variable markov oracle.
\newblock In Philippe Pasquier, Arne Eigenfeldt, and Oliver Bown, editors, {\em
  Musical Metacreation, Papers from the 2014 {AIIDE} Workshop, October 4, 2014,
  Raleigh, NC, {USA}}, volume {WS-14-18} of {\em {AAAI} Workshops}. {AAAI}
  Press, 2014.

\bibitem{tonicnet}
Omar Peracha.
\newblock Improving polyphonic music models with feature-rich encoding.
\newblock {\em CoRR}, abs/1911.11775, 2019.

\bibitem{crnngan}
Olof Mogren.
\newblock {C-RNN-GAN:} continuous recurrent neural networks with adversarial
  training.
\newblock {\em CoRR}, abs/1611.09904, 2016.

\bibitem{BrunnerWWZ18}
Gino Brunner, Yuyi Wang, Roger Wattenhofer, and Sumu Zhao.
\newblock Symbolic music genre transfer with cyclegan.
\newblock In Lefteri~H. Tsoukalas, {\'{E}}ric Gr{\'{e}}goire, and Miltiadis
  Alamaniotis, editors, {\em {IEEE} 30th International Conference on Tools with
  Artificial Intelligence, {ICTAI} 2018, 5-7 November 2018, Volos, Greece},
  pages 786--793. {IEEE}, 2018.

\bibitem{sternberg2018nature}
Robert~J Sternberg and James~C Kaufman.
\newblock {\em The nature of human creativity}.
\newblock Cambridge University Press, 2018.

\bibitem{midime}
Monica Dinculescu, Jesse Engel, and Adam Roberts, editors.
\newblock {\em MidiMe: Personalizing a MusicVAE model with user data}, 2019.

\end{thebibliography}

\end{document}